\documentclass[letter]{aa}  
\usepackage{fixltx2e}

\UseRawInputEncoding
\usepackage{float}
\usepackage{subfig}

\usepackage{graphicx}
\usepackage{txfonts}
\usepackage[bottom]{footmisc}
\usepackage{amsmath,bm}
\usepackage{xcolor}
\definecolor{xlinkcolor}{cmyk}{1,1,0,0}
\usepackage[utf8]{inputenc} % usually not needed (loaded by default)
\usepackage[T1]{fontenc}

\PassOptionsToPackage{hyphens}{url}
 \usepackage[bookmarks=true,         % show bookmarks bar?/ Changed March 22, 2019 for
     pdfnewwindow=true,      % links in new window
     colorlinks=true,    % false: boxed links; true: colored links
     linkcolor=xlinkcolor,     % color of internal links
     citecolor=xlinkcolor,     % color of links to bibliography
     filecolor=xlinkcolor,  % color of file links
     urlcolor=xlinkcolor,      % color of external links
     final=true,
 ]{hyperref}
\begin{document} 

   \title{Evidence of a vertical kinematic oscillation beyond the  \\Radcliffe Wave}
  % vertical oscillation is better?
   \subtitle{}

   \author{L.Thulasidharan\inst{1}\and %\fnmsep\thanks{}
          E. D'Onghia\inst{1,2} \and E. Poggio\inst{3,4} \and  R. Drimmel\inst{4} \and J. S. Gallagher III\inst{2} \and C. Swiggum\inst{5} \and \\ R. A. Benjamin\inst{6}\and  J. Alves\inst{5}
          }
  
   \institute{Department of Physics, University of Wisconsin-Madison, 1150 University Ave, Madison, WI 53706, USA\\
              \email{lthulasidhar@wisc.edu}
         \and
              Department of Astronomy, University of Wisconsin-Madison, 475 North Charter Street, Madison, WI 53706, USA
         \and
              Universit\'e C\`ote d'Azur, Observatoire de la C\`ote d'Azur, CNRS, Laboratoire Lagrange, Nice, France
         \and
             Osservatorio Astrofisico di Torino, Istituto Nazionale di Astrofisica (INAF), I-10025 Pino Torinese, Italy
        \and
             University of Vienna, Department of Astrophysics, T\"uurkenschanzstra{\ss}e 17, 1180 Vienna, Austria
        \and
            University of Wisconsin-Whitewater, Department of Physics, 800 West Main St, Whitewater, WI, 53190, USA
             }

   %\date{Received XXX; accepted XXX}

% \abstract{}{}{}{}{} 
% 5 {} token are mandatory

  \abstract{
%The RW is a 2.7-kiloparsec long
%  , damped sinusoidal wave-like arrangement of dense gas in the solar neighbourhood recently discovered by \citep{Alves2020}.
%wave-like feature in the kinematics?

The Radcliffe Wave (RW) is a recently discovered sinusoidal vertical feature of dense gas in the proximity of the Sun. In the disk plane, it is aligned with the Local Arm. However, the origin of its vertical undulation is still unknown. This study constrains the kinematics of the RW, using young stars and open clusters as tracers, and explores the possibility of this oscillation being part of a more extended vertical mode. We study the median vertical velocity trends of the young stars and clusters along with the RW and extend it further to the region beyond it. We discover a kinematic wave in the Galaxy, %-like kinematic feature %extending radially beyond the Radcliffe Wave 
distinct from the warp, with the amplitude of oscillation depending on the age of the stellar population. We perform a similar analysis in the N-body simulation of a satellite as massive as the Sagittarius dwarf galaxy impacting the galactic disk. %We predict a large-scale kinematic wave. %While the Radcliffe Wave and the oscillation beyond it may be interpreted as part of the local response of the disk to an external perturbation, 
When projected in the plane, the spiral density wave induced by the satellite impact is aligned with the RW, suggesting that both may be the response of the disk to an external perturbation.
However, the observed kinematic wave is misaligned. It appears as a kinematic wave travelling radially, winding up faster than the density wave matched by the RW,  
questioning its origin.
%However, the RW is misaligned with observed vertical kinematic wave  
%questioning its origin. 
If a satellite galaxy is responsible for this kinematic wave, we predict the existence of a vertical velocity dipole that should form across the disk and this may be measurable with the upcoming Gaia DR3 and DR4.}

   {}


   \maketitle

% Cam sentence: Recent revelations in astrometry have revealed the kinematics of the Radcliffe Wave's individual star-forming regions (CITE A BUNCH OF PAPERS HERE: Kounkel et al. 2018; Zari et al. 2019; Galli et al. 2019; Kuhn et al. 2019, 2020; Pavlidou et al. 2021; Großschedl et al. 2021; Swiggum et al. 2021). However, a dynamical connection between the regions along the Wave remains elusive and the nature and origin of this feature found in the dense gas is not well understood.

\section{Introduction}
%\textcolor{blue}{JSG comments}
%\textcolor{red}{Elena comments}
%\textcolor{green}{Lekshmi comments}
The Radcliffe Wave (RW) is a spatially and kinematically coherent, 2.7 kpc long arrangement of dense gas in the solar neighbourhood that was recently discovered by \cite{Alves2020}, in the work hereafter denoted as A20. It can be traced by the young star forming regions such as Orion, Perseus, Taurus, Cepheus and Cygnus. Using the precise distance measurements from \cite{zucker20}, it was found that the molecular clouds in this region align themselves to form a damped sinusoidal wave-like structure with a maximum vertical displacement of around 160 pc. Recent revelations in astrometry have unveiled the kinematics of individual star-forming regions of the RW \citep{kounkel2018, galli2019, kuhn2019, kuhn2020, zari19, pav2021, gro21, swiggum21}. However, a dynamical connection between the regions along the wave remains elusive and the nature and origin of this feature found in the dense gas are not well understood. Studies have shown similar distortions in the vertical direction for the nearby disk-galaxies \citep{edelsohn1997, matt2008, narayan20}, which are likely generated by a perturbation such as the passage of a satellite galaxy. Calculations also point to the possibility of the RW resulting from a Kelvin-Helmholtz instability \citep{Fleck2020}. A corrugation with a similar sinusoidal wave-like morphology discovered towards the inner galaxy, the Gangotri Wave, might be a potential RW sibling \citep{veena2021}. Another recently discovered corrugation feature in the solar neighbourhood traced by the OB stars, the Cepheus spur, may be related to the RW as  well \citep{pangon2021}. 

Our primary motivation behind this work is to identify the RW signature in the kinematics of young stars and clusters, as well as to understand the nature of this undulation using the vertical component of velocity. Thanks to the Gaia mission \citep{gaia16,gaia18a,gaia21}, we now have parallax and proper motion measurements for over a billion stars, which have helped in tracing the spiral arms in the solar neighbourhood using dynamically cold stellar populations  \citep{cantat18, cantat20, khop2019, kounkel2020, xu20, xu21, poggio21, zari21, medina21, castro21} and also to map out the Galactic warp in different stellar populations based on their age \citep{poggio18, lopez19, rom2019,chengdong19, cheng20, cantat20}.  

Evidence of north-south asymmetry in number density and bulk velocity of stars \citep{Widrow2012,BennetBovy2018} along with the recent discovery of phase-space spiral in the solar neighbourhood \citep{ant18} have established that the Milky Way (MW) disk is out of equilibrium, possibly due to the Sagittarius dwarf galaxy (Sgr) impact \citep{gom2013,bsch2018,bh2019,lapo2019b,bhtg2021,Hunt21,poggio21b}. In this work, we test, for the first time, the possibility of RW being a part of a large scale vertical mode (not the warp), using the kinematics of young stars and clusters. We then qualitatively compare our observations with the results from an N-body simulation to explore the scenario of this oscillation being caused by the impact of a satellite as massive as the Sgr in the Galactic disk.

Our paper is organised as follows. In Section 2, we describe the data samples used in our analysis, followed by our results in Section 3. Finally, in Section 4, we summarise and discuss our results and the open questions.
\vspace{-3mm}

\section{Data sample}

\begin{figure}
    \vspace{+3mm}
    \includegraphics[width=9cm]{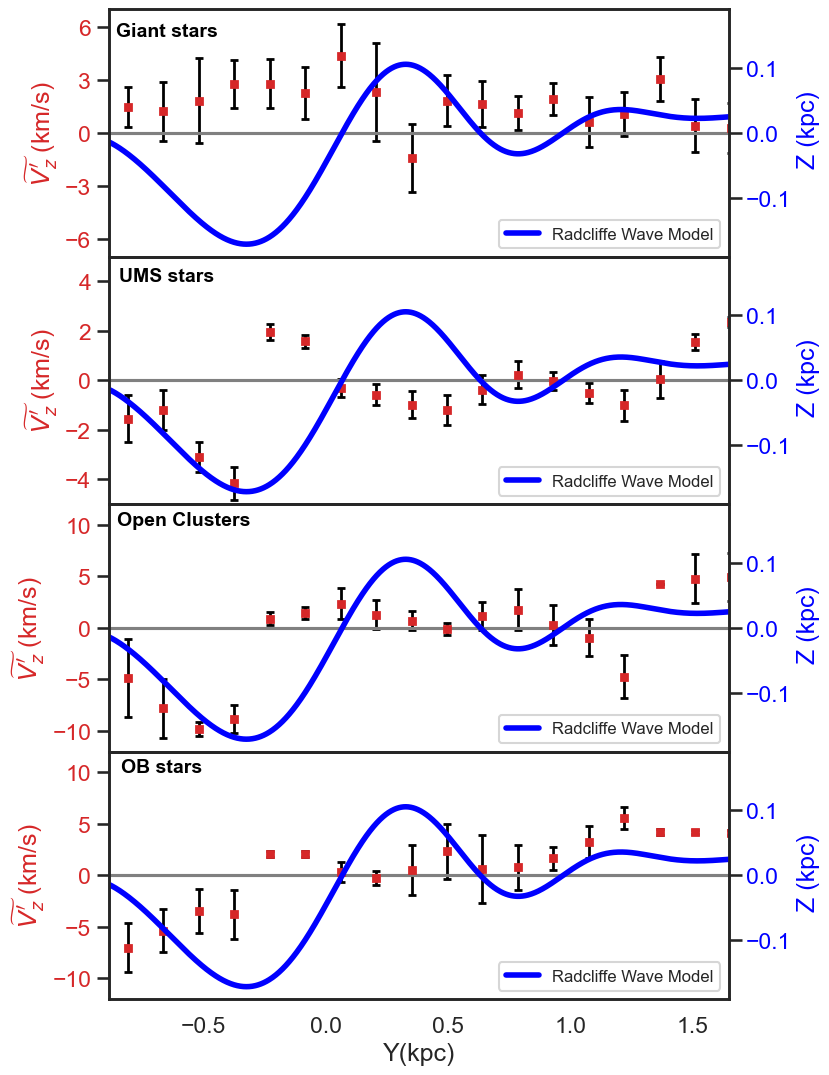}
    \caption{Kinematics of stars along the RW. The dual axis plot shows the position fit of RW model (blue curve) from A20 compared against the median vertical velocities of stars (red markers) in the respective overlapping bins of width 0.28 kpc. The vertical axes are coloured the same as that of their corresponding data points. The black solid lines represent the bootstrap uncertainties of the median.}
    \label{Radwave}
\end{figure}

In this section, we give an overview of the datasets that we used for this analysis. As shown in A20, molecular clouds are the best tracers of the RW. Thus, to constrain RW kinematics, we should ideally use the vertical velocities of these clouds, which are difficult to obtain due to the lack of proper motions. However, since this vertical oscillation is a feature mapped by the dense gas in the solar neighbourhood, we expect the young stars and clusters closer to this region to inherit the kinematic features of the RW. In contrast, older stars are dynamically more relaxed and, hence, they are not responsive to gas perturbations in the disk potential. Therefore, they are unlikely to show the signature of such vertical oscillations in their kinematics. For our analysis, we have taken three different young star samples available in the literature, namely the Open Cluster Sample \citep{cantat20} and the Alma catalogue of OB stars II \citep{pangon2021}, both of which are based on Gaia-DR2 data, as well as the Upper Main Sequence (UMS) sample consisting of OB and A stars, with updated astrometry from Gaia-EDR3 \citep{poggio21}. We cross-matched the UMS sample with the Alma catalogue to remove the common OB stars from the former. In addition to this, as a fiducial dataset, we use a sample of an older star population, the red giants from \cite{poggio18}, updated with Gaia-EDR3 astrometry. It may be worth mentioning here that the pre-main sequence stars from \cite{zari18}, as they are more closely associated with gas, may be better tracers for RW compared to the young stellar samples considered here. However, the sample only covered a portion of the RW from the first trough to first crest (blue curve in Figure \ref{Radwave}), not the entire range of RW.

\par
We selected the stars and clusters with $\varpi$/$\sigma_{\varpi}$ $>5$ and then we converted the sky coordinates and distances of these sources to 3D Galactic cartesian coordinates ($X, Y, Z$) using the Astropy package in python \citep{astropy18}. Since most of the young stars lack line-of-sight velocities, we derive their vertical velocity $V_z^{'}$ from proper motions, after correcting for differential Galactic rotation and solar motion using Eq. 8 in \cite{drimmel2000}. This adopted approximation for the vertical velocity is only valid for stars which are close to the Galactic midplane, where the contribution of the line-of-sight velocity component of the stellar peculiar motion to the vertical velocity is very small. We assume the recent rotation curve from \cite{eilers2019} with $V_c=229$ km/s, the components of Solar motion $(U_{\odot}, V_{\odot}, W_{\odot}) = (11.1, 12.24 , 7.25)$ km/s from \cite{schro2010} and the Galactocentric radius of the Sun $R_{0}=8.15$ kpc as in \cite{reid2019}.

We selected young stars with $|b|\leq20^\circ$, based on the aforementioned explanation, and we removed those with $R<5$ kpc, as the assumed rotation curve is not valid in this range of Galactocentric radii. In addition, for the Open Cluster sample, we applied an age cut at $100$ Myr to ensure the selection of only the young clusters. We note that for giants, we only selected the subset of stars that have line-of-sight velocity measurements available, so that full 3D space velocities can be derived. Our final sample consists of 598,446 UMS stars, 12,398 Alma OB stars, 620 open clusters, and 3,039,875 giants.

\begin{figure*}[htp]
    \centering
    \includegraphics[width=18.7cm]{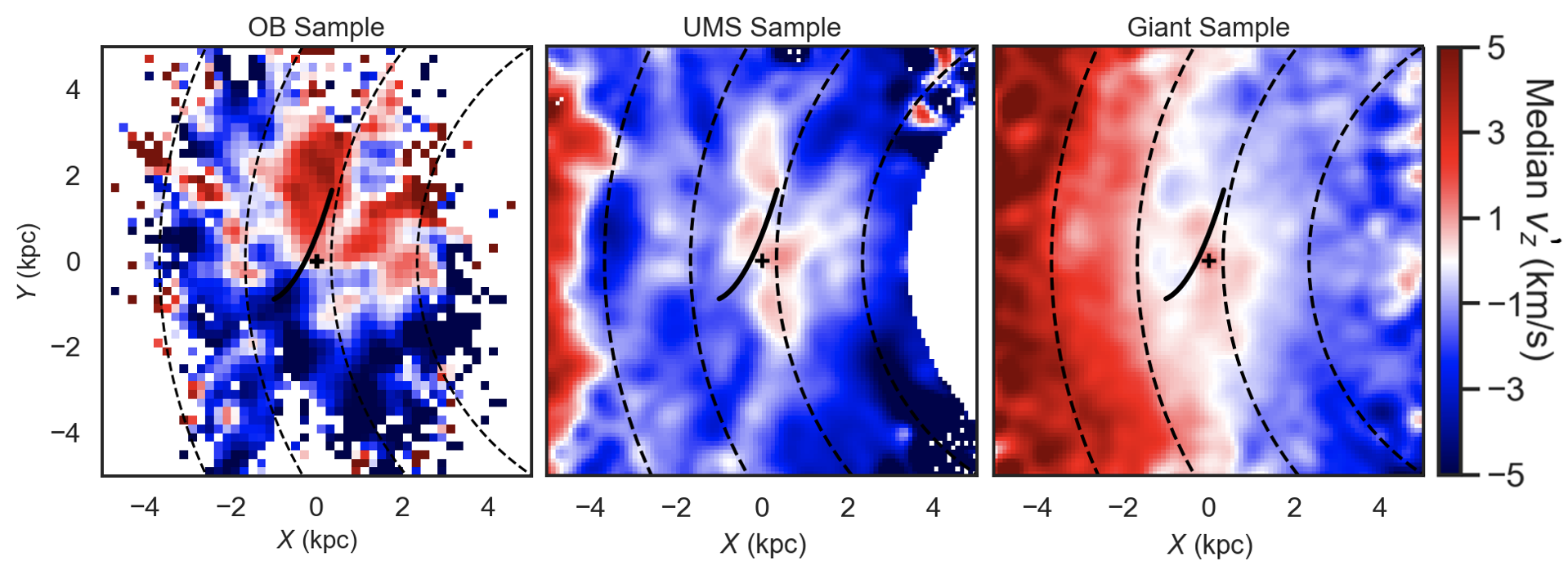}
    \includegraphics[width=18.7cm]{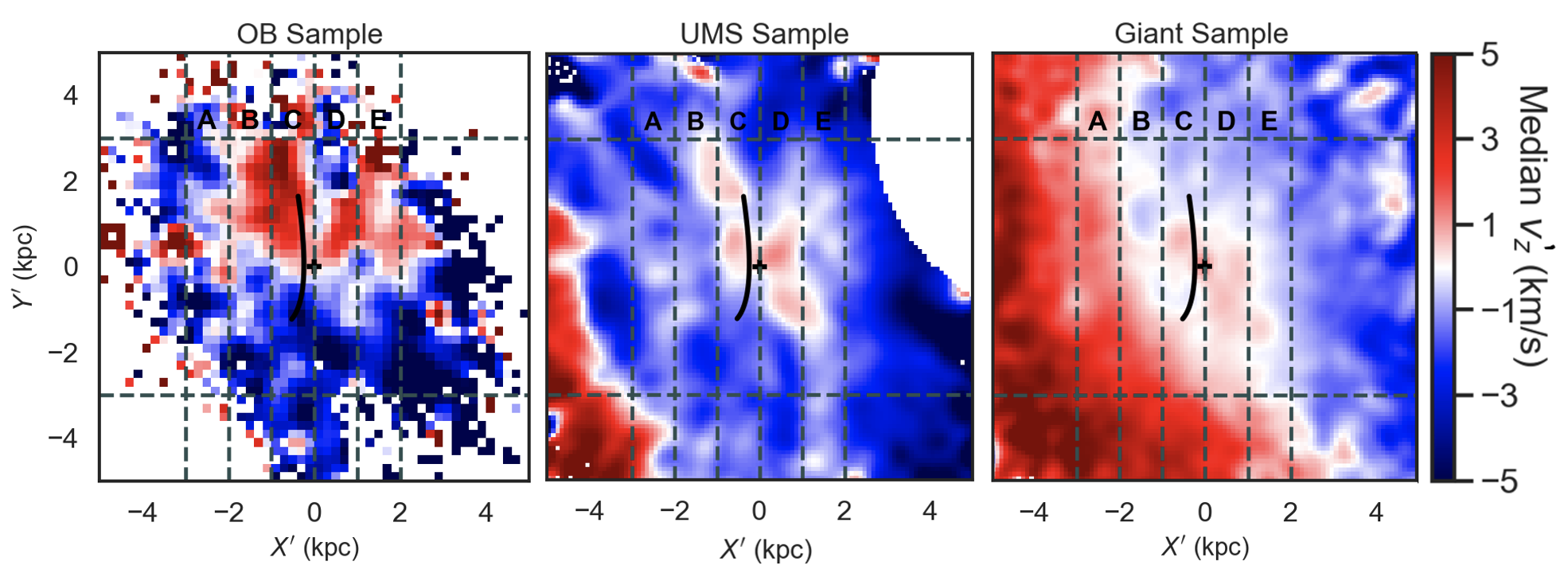}
    \caption{Vertical velocity maps of our samples showing a perturbation that extends beyond the RW (\textit{See text for details}). \textit{Top Row:} Cross at the centre shows the position of the sun, the black curve is the RW and black dashed curves from right to left corresponds to different Galactocentric radii from 6 to 12 kpc. \textit{Bottom Row:} Rotated coordinate system divided into five panels with panel C containing the RW.}
    \label{vel}
\end{figure*}

To study the kinematics of the RW, we chose those stars (both young and old) and the open clusters lying inside a cylindrical region that runs along the RW, with a radius of 80 pc and 160 pc, respectively. This choice of radius is motivated by the amplitude of the RW, however, changing the radius does not affect our results (see Fig \ref{vel}). This selection results in 2,862 UMS stars, 368 Alma OB stars, 54 open clusters, and 1,670 giants. We looked for signatures of vertical oscillations in the kinematics of young stars on a larger scale in the solar neighbourhood, beyond the RW. We then compared these signatures to those of the giants. 

\section{Results} 

%\begin{figure*}[!htp]
%    \centering
%    \includegraphics[width=19.2cm]{velocity_maps.png}
%\end{figure*}

\subsection{Constraining the kinematics of the RW}

Figure \ref{Radwave} displays the median vertical velocity ($\widetilde{V^{\prime}_{z}}$) distribution of different data samples selected along a cylinder that runs along the RW. The blue curve is a fit to the vertical displacement of the RW taken from A20 and we plot the $\widetilde{V^{\prime}_{z}}$ (shown in red) of young stars and clusters in overlapping bins of width 0.28 kpc along the RW and compare it against the red giants. The black solid lines represent the standard median of error calculated using bootstrap strategy.

We note that while the young stars and clusters show an oscillation in $\widetilde{ V^{`}_{z}}$ distribution, this trend is not seen in the sample of giant stars. Unlike the giant sample, young stars exhibit statistically significant deviations (larger than 3-$\sigma$) from $\widetilde{V^{`}_{z}}=0$ km/s (black solid line in Fig. 1). A consistent wave-like behaviour is apparent in the UMS, OB, and open cluster samples, particularly in the region starting from the first trough to the first crest of the RW, namely, from around -0.33 kpc to 0.33 kpc in $Y$. To demonstrate the damped wave-like behaviour mathematically, we  fit a simple 1-D harmonic oscillator model to the $\widetilde{ V^{`}_{z}}$ for UMS stars and Open Clusters (See Fig. \ref{fit} in the  Appendix). We note that the maximum amplitude in $\widetilde{ V^{`}_{z}}$ depends on the age of the star population, with the open clusters and OB sample showing a more prominent oscillation. To ensure that our cylindrical selection of stars along the RW do not force a wave-like feature in the kinematics, in addition to comparing the velocity features in young stars with the red giants, we placed the RW at random locations and random orientations in the disk and carried out the same analysis for different samples. We didn't observe similar consistent wave-like features in the kinematics of the stars at other locations, meaning that this undulation in the solar neighbourhood is real. In addition to this, using the Pingouin package in python \citep{vallat18}, we perform the Pearson rank correlation test on the  $\widetilde{ V^{`}_{z}}$ distribution of different stellar populations along the RW. We find a strong correlation between OB, UMS and the open cluster samples (OB-UMS with $p$-value$=0.0025$, UMS-Open cluster with $p$-value$=0.000009$ and OB-Open cluster with $p$-value$=0.00014$). However, no significant relationship has been noted among the young stellar samples and the giants (p-value $\sim$ 0.6-0.8).

If the disk's self-gravity were found to be driving the RW, we would expect a phase shift in the velocity by $90^\circ$ with respect to the position, as in the case of a simple harmonic oscillator. By taking the derivative of the RW model from A20, we can predict where the peaks and troughs should ideally be. However, the observed $\widetilde{ V^{`}_{z}}$ distribution is out of phase with such a distribution, as the peaks and dips in the data deviates from the expected, with a phase difference corresponding to around 0.16 kpc in position towards the left. This motivated us to extend our analysis beyond the region of RW, to look for signatures of vertical oscillations in the kinematics of young stars and to explore the possibility of the RW being a part of a more global vertical oscillation in the MW disk. 

\subsection{Kinematic maps}
We compared the $V_{z}^{\prime}$ distribution of different data samples within a cylinder of radius 3 kpc, with the Sun at the centre and the Galactic centre to the right, as shown in Figure \ref{vel} (top row). We divided the $X-Y$ plane into bins of width 100 pc for the UMS sample, 200 pc for the OB sample and 100 pc for the Giant sample, and we determined the median  $V^{\prime}_{z}$ of each bin. We also applied a simple Gaussian filter of kernel size 17x17 px for UMS and giant stars, and a filter of size 9x9 px for OB stars, to reduce noise and to better illustrate the vertical velocity trends in the data.

The young star samples show comparable distribution in  $\widetilde{ V^{\prime}_{z}}$, with the OB sample being the most perturbed. Most of the features seen in the OB stars seem to be washed out in the UMS sample as it contains the A-type stars. The Giant sample, consisting of older stars, does not show any significant signature of perturbations, as it is dominated by the Galactic warp starting within 2 kpc radius from the Sun. Due to the limited statistics, we are not showing the vertical velocity map of the Open Clusters, although the features on a larger scale are similar to those of the OB sample, which is quite expected. The velocity maps show that in young stars, there may be an oscillation in the vertical direction on a much larger scale than the RW (and even going beyond it), but is not part of the systematic warp signature. This perturbation is more evident in the youngest star population, which is dynamically colder and is thus more responsive, which, in our case, is represented by the OB sample. 

\subsection{Beyond the RW}
The RW appears as a nearly straight line when projected in the $X-Y$ plane, which makes it convenient to define a coordinate system aligned with this feature. We rotated the coordinate system $25^\circ$ anticlockwise with respect to the $Z$ axis, so that the rotated $Y$ axis (hereafter $Y^\prime$) lies tangential and the rotated $X$ axis (hereafter $X^\prime$) lies perpendicular to the RW. We then divided each of our samples into five panels A, B, C, D, and E starting from the left as in Figure \ref{vel} (bottom row), with panel C centred on the RW. Each panel is a rectangular region with $1$ kpc in width and $3$ kpc in length. We divided $Y^{\prime}$ into different bins and then calculate $\widetilde{ V^{`}_{z}}$ in each bin to look for oscillations beyond the RW. 
\begin{figure}
    \centering
    \includegraphics[width=9cm]{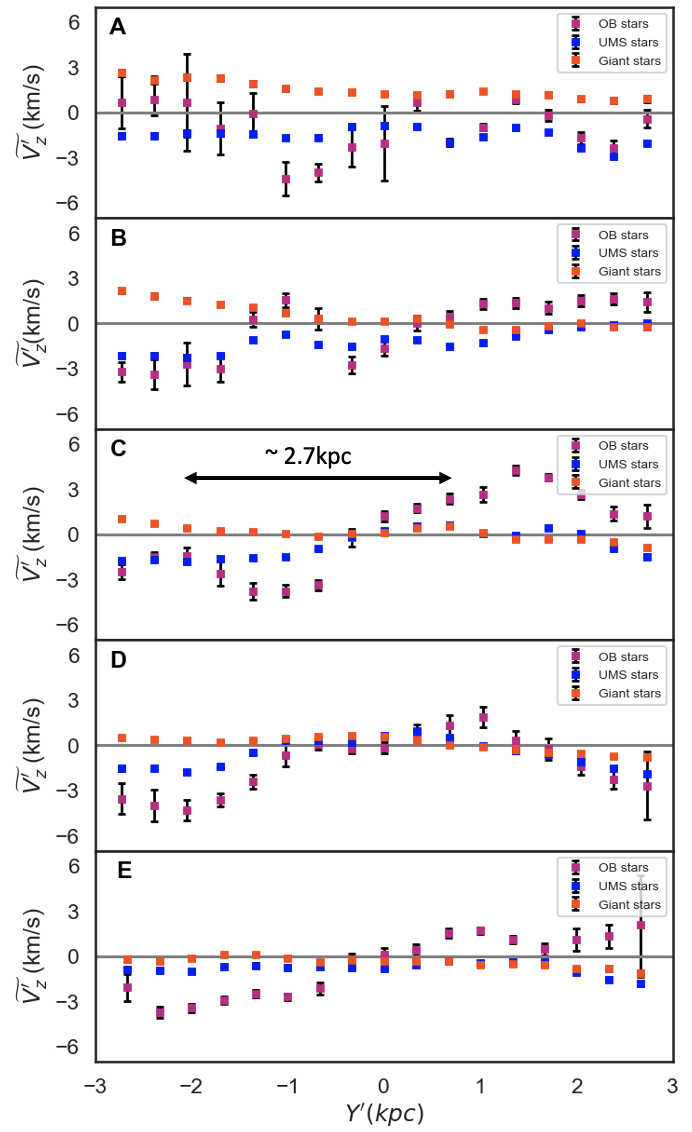}
    \caption{Age gradient in the oscillation of stars with the OB stars (youngest) showing a larger amplitude of oscillation in the kinematics. Each plot from top to bottom corresponds to the panels A to E as shown in the bottom row of Figure \ref{vel}. Here $Y^\prime$ is the coordinate along the Radcliffe Wave, as explained  in the text. The solid black lines correspond to the bootstrap uncertainties of $\widetilde{V_{z}^{\prime}}$.}
    \vspace{-2mm}
    \label{age}
\end{figure}

\begin{figure*}[!htp]
    \centering
    \includegraphics[width=18 cm]{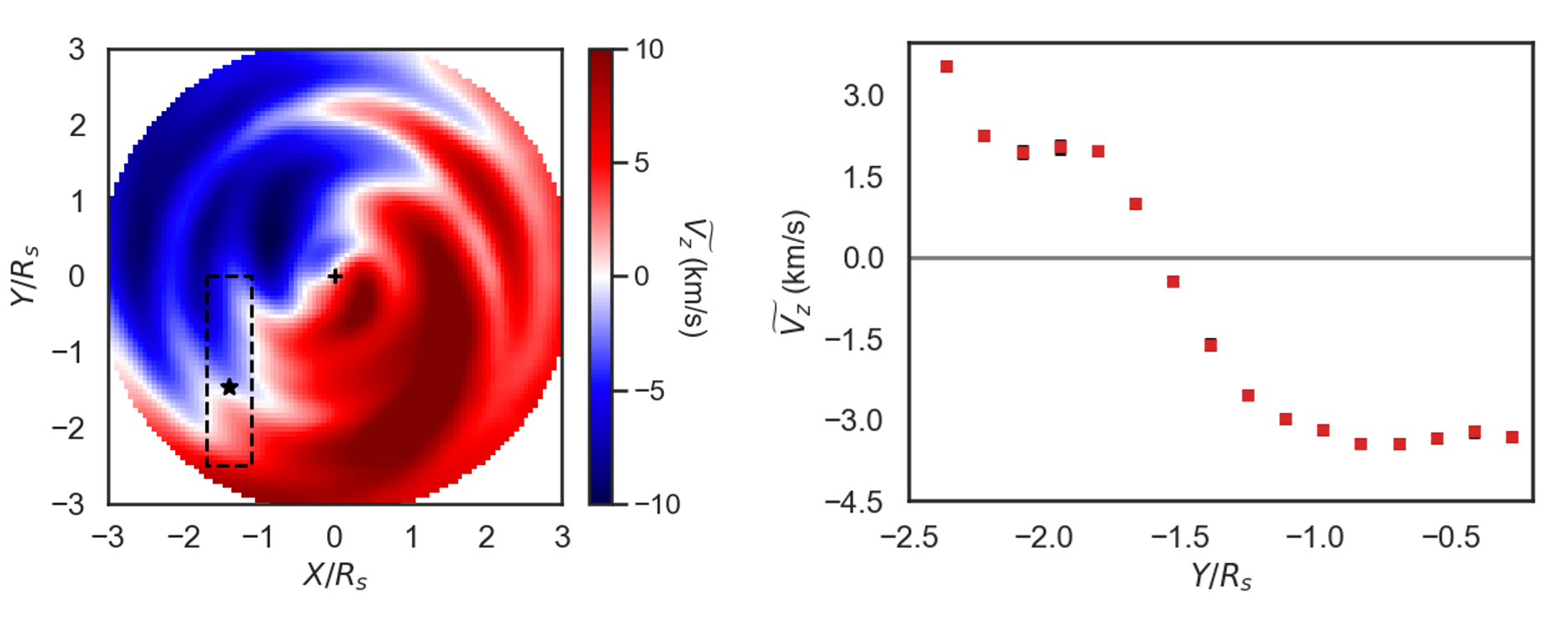}
    \caption{Median $V_z$ distribution of disk stars (a) projected in the $X-Y$ plane from the N-body simulation of a galactic disk perturbed by a satellite comparable in mass to the Sagittarius dwarf galaxy. (b) from the dashed rectangular region of the left plot, with the Sun placed at the interface of underdense (blue) and overdense (red) region in velocity space (as seen from the velocity maps in Figure 2), showing a maximum amplitude close to 4 km/s, comparable to that of the OB stars as seen from Figure \ref{vel}.} %The scale length of the disk $R_s$=4.83 kpc}
    \label{sim_vel}
\end{figure*}

\begin{figure*}[!htb]
    \centering
    \includegraphics[width=19cm]{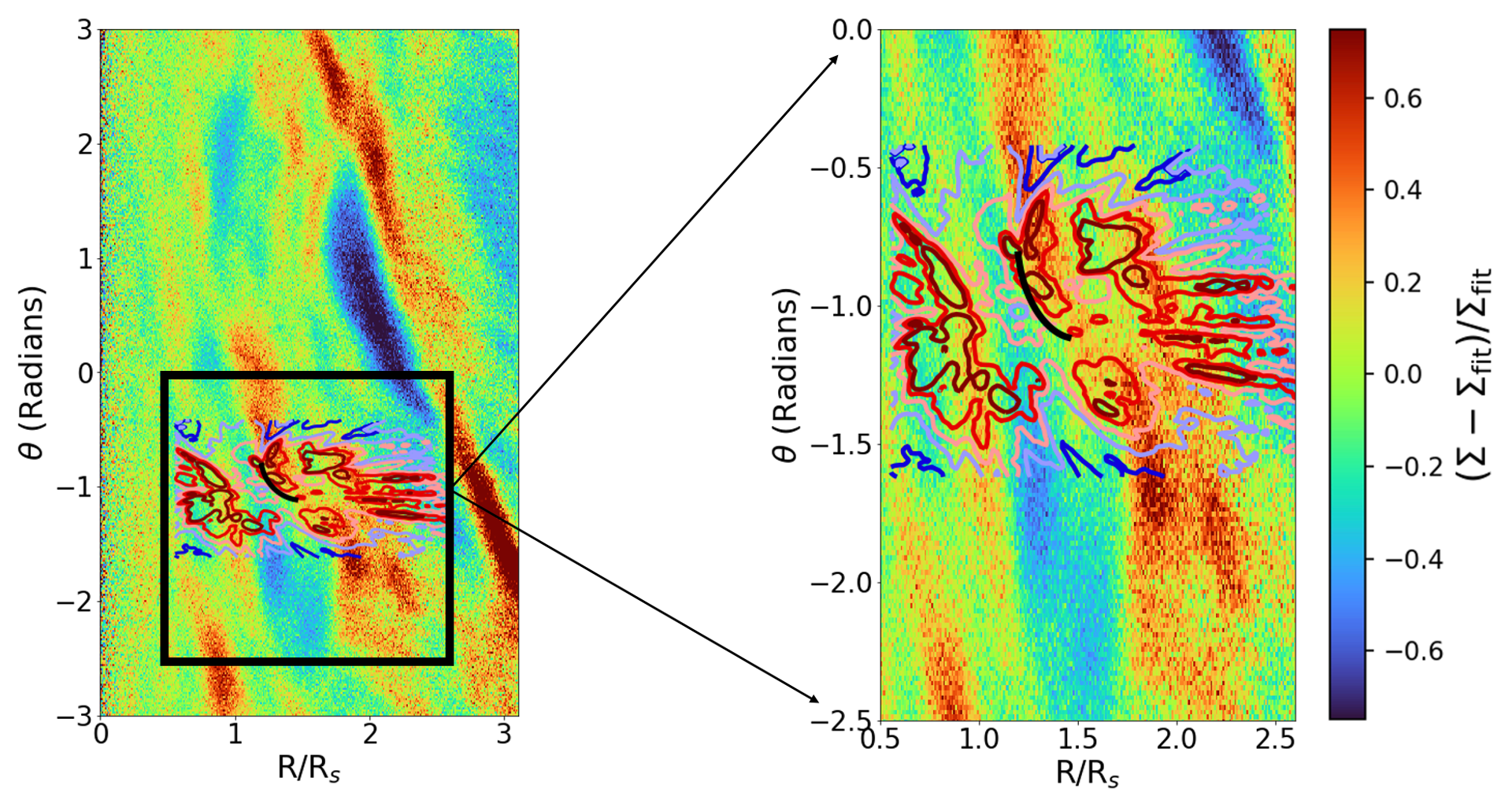}
    \caption{Overdensity contours of UMS data along with the Radcliffe Wave (black curve) overlapped with the simulation of a satellite galaxy perturbing the disk. The residual features are observed after t=0.6 Gyr of satellite punching the disk.The surface density residuals are computed as $\Sigma-\Sigma_{fit}/\Sigma_{fit}$ and displayed in polar coordinates.}
    \label{density}
\end{figure*}
 
The results are summarised in Figure \ref{age}. This plot shows the existence of a vertical kinematic oscillation in the region parallel to the RW, which appears to be `travelling' radially towards the Galactic anti-centre .The wavelength of this kinematic wave is around $2.5-2.7$ kpc and the maximum oscillation is seen in the panel C, which contains the RW and panel D. We also observe an age gradient in the oscillation of stars in each panel, with the youngest sample (here, OB stars) showing a maximum response to the perturbation. The maximum velocity amplitude of the wave in the OB sample is around $3.77$ km/s in panel C and $4.28$ km/s in panel D, compared to $1.49$ km/s in C and $1.78$ km/s in D in the UMS sample, at $Y^{\prime}=-1.33$ kpc and $-2.0$ kpc, respectively. We also performed the Pearson rank correlation test between the median vertical velocities of the three stellar samples and found a strong association between the undulations seen in OB and UMS stars (for e.g., p-value of 0.0002 in panel C). At the same time, OB-Giant and UMS-Giant combinations do not show any significant correlation in any of the panels (for e.g., p-value of 0.13 and 0.52, respectively, for panel C).

%\textcolor{blue}{Useful to briefly summarize the properties of the n-body simulation for the reader, including satellite mass and angle of interaction, bound orbit, etc.} %clarified

We performed the same analysis in the N-body simulation of a galactic disk perturbed by a satellite as massive as the Sgr. We added a live substructure to the stellar disk, with the total mass of $2\times10^{10}$ M$_{\odot}$ which hits the disk at roughly 2.2 kpc from the galactic centre. The initial position of the satellite was at 20 kpc with an eccentric orbit ($e=0.7$). It completed one orbit around the Galactic centre over approximately 1.7 Gyrs. In the numerical experiment, the satellite excites the stellar disk, as seen by the formation of a long-wavelength bending mode \citep{donghia2016}. 

Figure \ref{sim_vel} displays the median vertical velocity of the simulated disk stars projected in the X-Y plane after $\sim$0.6 Gyr of evolution past the impact (left panel). The interaction with the satellite causes a dipole in the vertical velocity. When the Sun (marked with the black star) is located in an underdense region of the disk, we note that the predicted velocity map of the simulation (dashed rectangular area) qualitatively matches the observed one (left panel of Figure \ref{vel}). Also, the disk oscillates with an amplitude of 4-5 km/s (right panel of Figure \ref{sim_vel}), comparable to the observed OB stars in the experimental data (Figure \ref{age}). The wavelength of the bending mode in the simulation is around 10 kpc, which is greater than the disk scale length ($R_s$=4 kpc) and is much greater than the wavelength of the kinematic vertical oscillation observed in the data. We note that the spatial and mass resolution of the simulation are insufficient to resolve the region observed in the current data in the phase space and hence, to describe the response of the perturbation at the scale of the observations reported here. The orbital configuration of the satellite-MW disk does not represent the interaction with Sagittarius but a more general interaction with a satellite impacting the disk and the simulation does not include the gas component. However, the velocity dipole in the median of the stellar vertical velocity is expected to form similarly as a general case and may be observable with the upcoming Gaia DR3 and DR4. Figure \ref{density} shows the relative surface density residuals of the simulated disk after $\sim$0.6 Gyr of the evolution of the disk after the impact of the satellite on the galactic disk. We used the Fourier transform analysis to estimate the residuals of the surface density, computed by subtracting the azimuthally averaged surface density from the surface density distribution and normalising it according to $Res=\Sigma-\Sigma_{fit}/\Sigma_{fit}$. The overdensity contours of the UMS data, as mapped by \citet{poggio21}, are projected in the polar plane and overlap the simulated disk and enclose the observed RW (solid black line). We note that the pitch angle of the RW qualitatively agrees in density with the spiral structure induced in the simulation by the satellite impact. While this outcome supports the evidence that the RW (or, at least, its stellar counterpart) may be part of the disk response to the perturbation induced by a satellite galaxy, Fig. 2 shows that, in terms of kinematics, the RW and the large-scale vertical oscillation are not aligned, calling into question the assumption of their having a common origin. 

\section{Discussion and conclusions}
The existence of the RW as a vertical undulation that connects the nearby known star-forming regions is now well established \citep[but see][for a different view]{donada21}. We find that the young stars associated with the RW show a wave-like pattern in their kinematics, suggesting that the vertical oscillation of RW indeed has a kinematic origin.
%Using a sample of OB stars, UMS stars, and young open clusters as tracers of the RW, we studied its kinematics, inferred velocity maps of the young stars beyond it and compared it against red giants.
In particular, we showed that the stellar counterpart of the RW relates to a more extended vertical oscillation measurable across a larger radius from $X^{\prime}=2$ kpc to $-3$ kpc in a frame centred on the Sun. Our analysis shows that this sinusoidal feature, which extends radially beyond the RW, has an amplitude of oscillation that depends on the age of the stellar population. The dynamically cool younger stars are taking part, whereas the dynamically warmer older stars are not responsive to the perturbation. These findings bring the nature of this vertical feature into question.  

A recent study suggested that the RW might be due to the Kelvin-Helmholtz instability \citep{Fleck2020}. However, the vertical displacement of this oscillation is close to 200 pc above and below the disk mid-plane - a large extension that might suggest a different origin. Another feature possibly related to the RW is the Cepheus spur \citep{pangon2021}, a bridge connecting the Orion-Cygnus spiral arm with the arm of Perseus, near the vicinity of the Sun. It is possible that what we see here may be a continuation of that spur. However, the kinematic wave that we observe here is extended at greater radii than the Cepheus structure, indicating that we may be witnessing evidence of a corrugation in the disk.

Recent simulations have investigated the Large Magellanic Cloud (LMC) interaction with the MW. Those studies showed that in allowing the MW dark halo to be `alive' and responding actively to the potential, the LMC might produce a wake in the halo, which then distorts the disk  \citep{Weinberg1998, garavito2019, lucchini2021}. While the interaction of the Large Magellanic Cloud (LMC)  with the MW may have induced the formation of the gas and stellar warp \citep{binney1992,wb2006,poggio20}, the Sagittarius dwarf galaxy may also disturb the halo forming an overdensity in the satellite's proximity \citep{lapo2018b}. There are reasons to believe the halo perturbation induced by the dwarf galaxy may also have affected the disk, producing the oscillation we observe. The analysis of a numerical experiment with an MW-satellite interaction \citep{donghia2016} shows that a satellite impacting the disk may generate a wobble in the disk with an associated vertical velocity dipole. 

%However, we note that the wavelength in the simulation is greater than the wavelength observed in the current data (2.7 kpc).

%clarified \textcolor{blue}{Do we have a candidate satellite that could have been the impactor, or can we argue that it is long gone and so we don't see it? YES SAGITTARIUS}

%While the current data are limited in radial extension and thus cannot entirely prove the existence of this long-wave mode across the entire disk, 
We speculate that the undulation reported here may, in fact, be the local disk response to this external perturbation. We question whether the RW has taken part in this process. Because the gas is dynamically colder than the stellar disk, the RW and the young stars will be more responsive to the perturbation than the older stars in the disk, explaining the gradient in the amplitude of the oscillation observed in Figure \ref{age}, as well as the shorter wavelength. We note that the amplitude of the simulated kinematic vertical oscillation of a dynamically cold disk matches the observed behaviour of the stars in the region where the RW is located, suggesting a possible common origin. 

 While the pitch angle of the RW is similar to the simulated spiral structure formed by the impact of a satellite dwarf galaxy, which appears round, in the form of rings propagating outwards \citep[see e.g.][]{donghia2016,lapo2019b}, the kinematic wave beyond the RW has a different alignment. Recent simulations by \cite{bhtg2021} showed that $m=2$ density spiral waves induced by an external perturber wind up faster than the associated kinematic waves. Our analysis of the data points to the opposite conclusion. Figure \ref{vel} shows that the observed kinematic oscillation indeed has a different alignment than the RW, but with a smaller pitch angle, suggesting it winds up faster than the spiral density wave. Another possibility is that it is a radial wave, indicating that the picture is more complex, perhaps due to the superposition of different Fourier terms or to multiple passages of a satellite, or both \citep[see panels c and d of Fig. 12 in][]{poggio21b}. This evidence questions whether RW is a part of this disk response, or rather a local in-plane density perturbation `riding on top' of the wave-like kinematic perturbation induced by a satellite galaxy.  
 
 The upcoming Gaia DR3 and DR4 will confirm the existence of the vertical oscillations that we display in Figure \ref{age}, by inferring vertical velocity maps of stars across the disk and measuring its associated dipole. At the same time, this test may elucidate the nature of the RW.

\begin{acknowledgements}
    We thank  Michelangelo Pantaleoni Gonz{\'a}lez and Dr. Jes\'us Ma\'iz Apell\'aniz for providing the Alma catalogue of OB stars (II). ED thanks Martin Weinberg for insightful discussions. We thank the anonymous referee whose comments helped improve and clarify this manuscript.

\end{acknowledgements}

\begin{appendix} %First appendix
%\onecolumn
\section{Damped sinusoidal model for Radcliffe Wave}

\begin{figure}[!htb]
    %\centering
    \includegraphics[width=8cm]{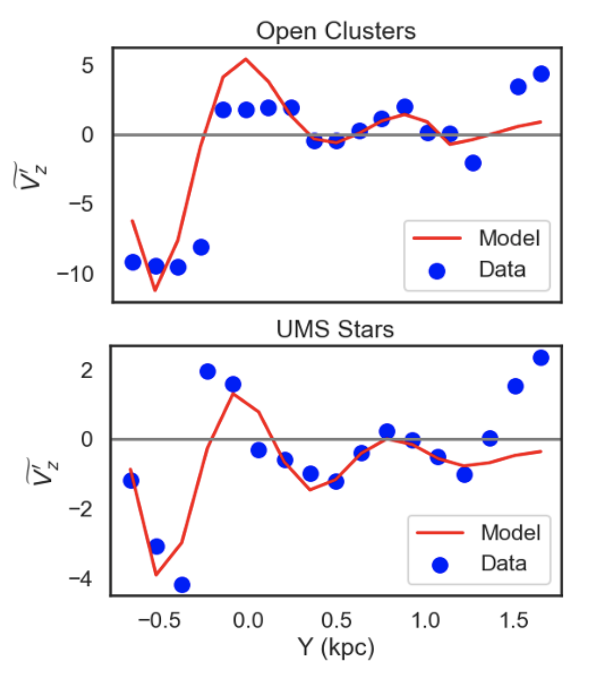}
    \caption{Fit for Open clusters and UMS stars assuming a 1D damped harmonic oscillator model where we define $Z(y)=A e^{-\gamma y} \sin{\omega y}$ and its derivative $V_{z}(y)=-A\gamma e^{-\gamma y} \sin{\omega y} + A\omega e^{-\gamma y} \cos{\omega y}$.} 
    \label{fit}
\end{figure}

In order to mathematically demonstrate the damped wave-like feature seen with different tracers along with the Radcliffe Wave as shown in Figure \ref{Radwave}, we fit a 1-D damped harmonic oscillator model to the $\widetilde{ V^{\prime}_{z}}$ (Figure \ref{fit}). We define $Z(y)=A e^{-\gamma y} \sin{\omega y}$ and its derivative $V_{z}(y)=-A\gamma e^{-\gamma y} \sin{\omega y} + A\omega e^{-\gamma y} \cos{\omega y}$, where $\gamma$ is the damping coefficient and $\omega$ is the oscillation frequency.

We studied the median vertical velocity behaviour and we do not have enough data points to produce a well-defined fit for the velocity profile. However, the $\widetilde{V_{z}^{\prime}}$  distribution follows a damped wave-like profile.

\end{appendix}
\end{document}